# Multi-sensor system for simultaneous ultra-low-field MRI and MEG

Vadim S. Zotev[a,*], Andrei N. Matlachov[a], Petr L. Volegov[a],

Henrik J. Sandin[a], Michelle A. Espy[a], John C. Mosher[a],

Algis V. Urbaitis[a], Shaun G. Newman[a], Robert H. Kraus, Jr[a].

[a]*Los Alamos National Laboratory, Group of Applied Modern Physics MS D454, Los Alamos, NM 87545, USA*

**Abstract.** Magnetoencephalography (MEG) and magnetic resonance imaging at ultra-low fields (ULF MRI) are two methods based on the ability of SQUID (superconducting quantum interference device) sensors to detect femtotesla magnetic fields. Combination of these methods will allow simultaneous functional (MEG) and structural (ULF MRI) imaging of the human brain. In this paper, we report the first implementation of a multi-sensor SQUID system designed for both MEG and ULF MRI. We present a multi-channel image of a human hand obtained at 46 microtesla field, as well as results of auditory MEG measurements with the new system.

*Keywords:* biomagnetism; MEG; MRI; SQUID; co-registration

## 1. Introduction

Magnetoencephalography (MEG), that measures magnetic fields directly associated with neuronal activity [1], is a well-established technique for studies of brain function [2]. The high temporal resolution and non-invasive nature of MEG make it a valuable tool for both fundamental brain research and clinical diagnostics such as pre-surgical evaluation in treatment of epilepsy [3]. In many applications requiring spatial information, magnetic sources, determined by the MEG, have to be precisely mapped onto anatomical brain structure, usually obtained with the conventional high-field magnetic resonance imaging (MRI). Therefore, spatiotemporal brain imaging based on the fusion of MEG and MRI data requires two very different (and expensive) instruments. Moreover, in addition to the MEG source localization errors [2], there are errors associated with MEG/MRI co-registration.

[*] Corresponding author. Tel: 1-505-665-8460. Fax: 1-505-665-4507.
*E-mail address*: vzotev@lanl.gov.

A typical MEG/MRI co-registration procedure includes three steps [3,4]. First, the MEG sensor array is used to detect positions of several small localization coils connected to the subject's head. Then, both the head surface and the localization coils are digitized by means of the Polhemus 3D digitizer. Finally, the digitized head surface is matched with the scalp surface obtained from the subject's MRI. This procedure, apart from being time-consuming and laborious, introduces errors. The Aston group recently reported average co-registration error of 4 mm when using localization coils mounted on a bite-bar, and 9 mm error in case of the coils attached to the scalp [4]. This level of inaccuracy (~5 mm and more) appears typical in MEG/MRI co-registration.

The best way to solve this problem is to eliminate the need for co-registration altogether by acquiring both MEG signals and anatomical images with the same sensors. This can be achieved if one uses the MEG sensor array to perform magnetic resonance imaging at ultra-low fields (ULF MRI). In this imaging method, introduced by the Berkeley group [5-7], the nuclear spin population in a sample is pre-polarized [8] by a relatively strong (up to 0.1 T) magnetic field, and the spin precession is encoded and detected at an ultra-low (microtesla range) measurement field after the pre-polarizing field is removed [5-8]. Because sensors of the same type – superconducting quantum interference device (SQUID) sensors – are used for both MEG [2] and ULF MRI [5], the two techniques are compatible. Moreover, it was demonstrated recently by our group that MEG and ULF MRI signals could be measured simultaneously with the same SQUID sensor and separated by filtering during data processing [9]. The combination of these two methods will allow simultaneous functional (MEG) and structural (ULF MRI) imaging of the brain with the additional benefits of reduced imaging time and cost.

In this paper, we report the first implementation of a multi-sensor SQUID system, specifically designed for measuring both MEG and ULF MRI signals, and present initial results demonstrating the system's performance.

## 2. Materials and Methods

Design of the system we have developed for measurements of MEG and ULF MRI has been described in detail elsewhere [10]. In this section, we only specify the main physical characteristics of the system and experimental parameters.

The system includes seven second-order gradiometers with SQUID sensors installed inside a liquid He dewar. The gradiometers have 37 mm diameter and 60 mm baseline. They are arranged in parallel (with one gradiometer in the middle and six others surrounding it in a circle) with 45 mm separations between coil centers. The noise spectral density is 1.2 fT/√Hz at 1 kHz for the central channel and 2.5-2.8 fT/√Hz for the surrounding channels. The noise spectra are essentially flat down to ~3 Hz.

The ULF MRI coil system is designed for 3-D Fourier imaging of large objects such as the human brain. However, only 2-D imaging experiments have been performed so far. In section 3, we present a multi-channel 2-D image of the human hand. It was obtained at the measurement field of 46 μT (frequency ~1950 Hz), with a pre-polarization field of 10 mT. The standard 2-D gradient echo imaging sequence was employed with the encoding time of 40 ms and acquisition time of 80 ms. The limiting values of the two gradients were ±94 μT/m (±40 Hz/cm), yielding 3 by 3 mm imaging resolution.

## 3. Results

Results of 2-D ULF MRI of the human hand are exhibited in Figs. 1 and 2. The hand was positioned under the bottom of the dewar. Fig. 1 shows images from the seven individual channels. The frequency and phase encoding directions are the horizontal and vertical directions in Fig. 1, respectively. The composite image, exhibited in Fig. 2, was obtained as a sum of the seven individual images, and then sensitivity-corrected with the use of an experimentally determined seven-channel sensitivity map.

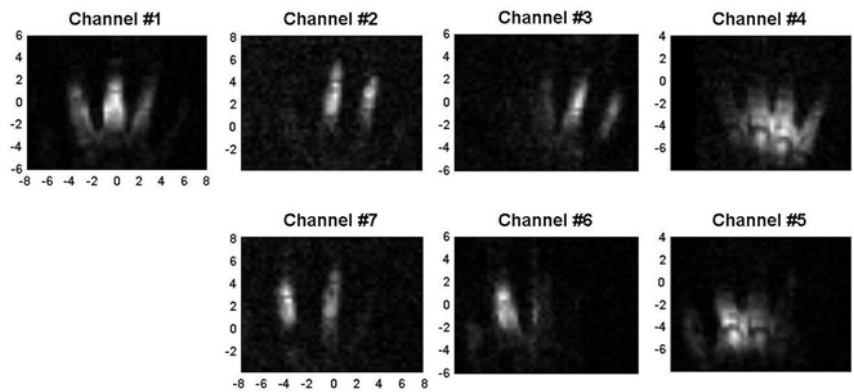

Fig. 1. Images of the human hand measured by the seven individual channels. The coordinates are in cm.

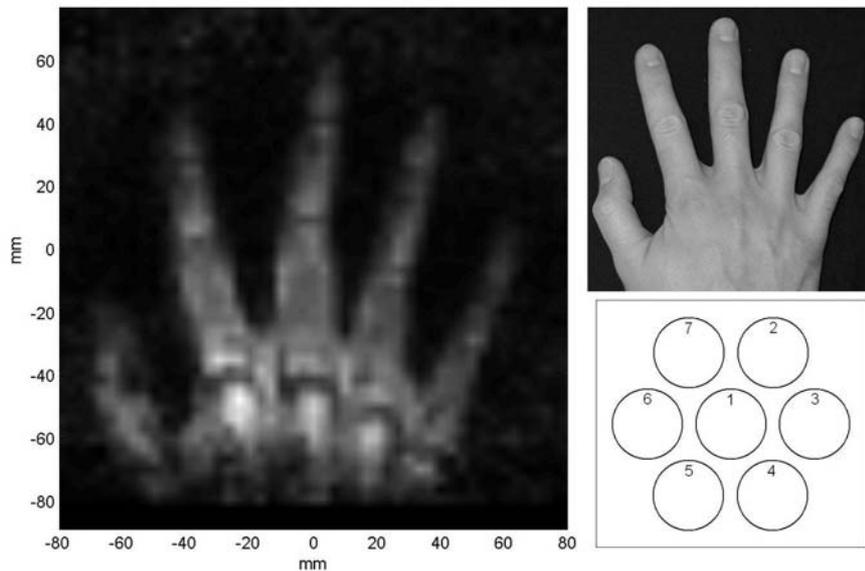

Fig. 2. *Left*: The composite image with sensitivity correction. *Right*: The subject's hand and the channel map.

To test the system's MEG performance, we carried out an auditory MEG experiment. The auditory stimulus (a train of 1.2 ms clicks following with 14 ms intervals) was applied 50 ms after the start of data acquisition. The auditory evoked magnetic fields, measured by the seven channels, are exhibited in Fig. 3. They are consistent with the orientation of the equivalent current dipole perpendicular to the Sylvian fissure [2].

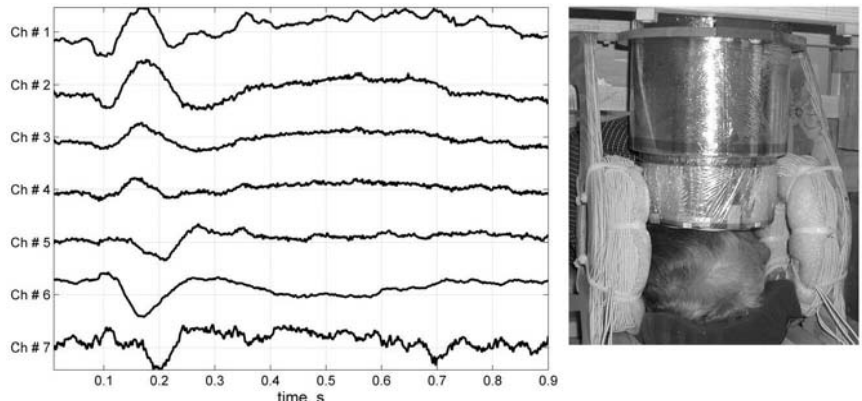

Fig. 3. *Left*: Auditory MEG measured by the seven channels. *Right*: Position of the head during the experiment.

## 4. Conclusion

The experimental results, presented in this work, demonstrate that our seven-channel SQUID system can be successfully used for both human-subject ULF MRI and MEG. Our next goal is to implement 3-D ULF MRI of the human brain and thus prove feasibility of simultaneous functional and anatomical brain imaging.

## References

[1] D. Cohen. Magnetoencephalography: evidence of magnetic fields produced by alpha-rhythm currents. Science 1968;161:784-786.
[2] M. Hämäläinen, R. Hari, R.J. Ilmoniemi, J. Knuutila, O.V. Lounasmaa. Magnetoencephalography – theory, instrumentation, and applications to non-invasive studies of the working human brain. Rev Mod Phys 1993;65:413-505.
[3] S. Knake, E. Halgren, H. Shiraishi, et al. The value of multichannel MEG and EEG in the presurgical evaluation of 70 epilepsy patients. Epilepsy Res 2006;69:80-86.
[4] P. Adjamian, G.R. Barnes, A. Hillebrand, et al. Co-registration of magnetoencephalography with magnetic resonance imaging using bite-bar-based fiducials and surface matching. Clin Neurophysiol 2004;115:691-698.
[5] R. McDermott, S.K. Lee, B. ten Haken, A.H. Trabesinger, A. Pines, J. Clarke. Microtesla MRI with a superconducting quantum interference device. Proc Nat Acad Sci 2004;101:7857-7861.
[6] R. McDermott, N. Kelso, S.K. Lee, et al. SQUID-detected magnetic resonance imaging in microtesla magnetic fields. J Low Temp Phys 2004;135:793-821.
[7] M. Mößle, W.R. Myers, S.K. Lee, et al. SQUID-detected in vivo MRI at microtesla magnetic fields. IEEE Trans Appl Superconduct 2005;15:757-760.
[8] A. Macovski, S. Conolly. Novel approaches to low cost MRI. Magn Reson Med 1993;30:221-230.
[9] P. Volegov, A.N. Matlachov, M.A. Espy, J.S. George, R.H. Kraus, Jr. Simultaneous magnetoencephalography and SQUID detected nuclear MR in microtesla magnetic fields. Magn Reson Med 2004;52:467-470.
[10] V.S. Zotev, A.N. Matlachov, P.L. Volegov, et al. Submitted to IEEE Trans Appl Superconduct.